\documentstyle[aps,prl,epsf]{revtex}
\begin{document}

\twocolumn[\hsize\textwidth\columnwidth\hsize\csname
@twocolumnfalse\endcsname

\vskip2pc] \narrowtext

{\bf Comment on ``First order amorphous-
amorphous transformation in silica''} 

In a recent letter\cite{sio2pa}, Lacks presents evidence of a first order
amorphous-amorphous transition in silica at $T=0$. He bases his conclusions
on the calculation of the free energy along a path of compression and successive
decompression of a sample of 108 SiO$_2$ units. The free energy of the two 
branches cross each other, and this is interpreted as evidence
of a first order transition, which is suggested to be hidden in the
actual simulations (and in experiments) due to dynamic arrest. 
We show that this conclusion  does not follow from the shown data, since qualitatively
the same phenomenology is obtained in a model where 
a first order transition does not exist. 

\begin{figure}
\narrowtext
\epsfxsize=3.3truein
\vbox{\hskip 0.05truein
\epsffile{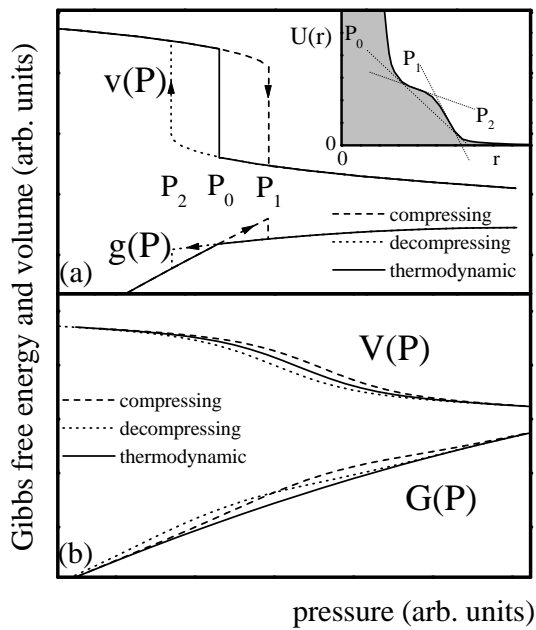}}
\medskip
\caption{(a) Free energy g(P) and volume v(P) forms for a 1D interaction potential 
with two equilibrium distances (sketched
in the inset). (b) The result of averaging over a disordered sample.}
\label{f1}
\end{figure}

The model we use 
has been applied to study the anomalous properties of
tetrahedrally coordinated liquids\cite{edu}, of which silica is an example. 
It consist of particles interacting through
a spherical, purely repulsive potential $U(r)$
as qualitatively depicted in the inset of Fig. 1. Its main property is the existence of two 
different equilibrium distances depending on pressure.
We will work out here the
one-dimensional (1D) case, that can be solved exactly, but we have performed numerical 
simulations in more realistic two- and three-dimensional systems, 
with the same qualitative results \cite{eduump} (see also \cite{edu}, Fig. 7).
For the 1D system at zero temperature, the qualitative evolution of Gibbs free
energy and volume vs. external pressure for a pair
of neighbor particles is given in Fig. 1(a). There is a bistability
upon compression and 
decompression, as indicated in the figure. 
When an assembly of particles
is considered, we have to sum over all of them to get the result for the whole system. 
In 1D, polydispersity should be
introduced to avoid `crystallization'.
The free energy has to be calculated as $G(P)=\sum_i  g^i(P)$, where $i$ labels the
particles, and $g^i(P)$
is the free energy of each particle, which is not the same throughout the system due to polydispersity
(in real glasses, the abundance of different local 
environments plays the role of the polydispersity we introduce in the 1D case).
Depending on what we are interested in, the compression, decompression, or thermodynamic 
forms of  $g^i(P)$ have to be used. An analog
expression holds for $V(P)$. 

The qualitative final forms of $G(P)$ and $V(P)$ are shown in Fig. 1(b).  The free energy of 
compression and decompression branches cross at some point (which is reminiscent of the 
microscopic $g(P)$ form) as in Fig. 2(a) of \cite{sio2pa}. 
However, the thermodynamic $G(P)$ is always lower, and smooth for all $P$,
without signs of a first order transition.
Then we conclude that the evidence presented in \cite{sio2pa} for silica 
is not neccesarely related to a first order
transition. The definite answer must be sought in simulations
in which the system is annealed at different $P$, to
obtain the thermodynamic form of $G(P)$ and $V(P)$. For $T>0$
we expect hysteresis in $V(P)$ and $G(P)$ to disappear in simulations 
(or experiments) performed at $T$
higher than the glass temperature $T_g$, at which dynamics freezes. 

Finishing, we notice that the presence of an attractive part in the 
interaction potential may produce (if strong enough) 
a first order amorphous-amorphous transition in the model\cite{edu}. 
In this case, numerical simulations at zero temperature in a
two dimensional system \cite{eduump}
show that our model reproduces qualitatively the $V(P)$ behavior observed 
in water \cite{mishima}. This points out that a first order
amorphous-amorphous transition--if it exists--can be directly observed in 
numerical simulations even well below $T_g$.
 
\vspace{0.5cm}
E. A. Jagla

{\small 
Centro At\'omico Bariloche,
(8400) Bariloche, Argentina}


\end{document}